\documentclass[aps,prc,twocolumn,nofootinbib,preprintnumbers,amsmath,amssymb]{revtex4}

\usepackage{graphicx}
\usepackage{dcolumn}
\usepackage{bm}
\usepackage{epsfig}
\usepackage{longtable}

\def\fun#1#2{\lower3.6pt\vbox{\baselineskip0pt\lineskip.9pt
\ialign{$\mathsurround=0pt#1\hfil##\hfil$\crcr#2\crcr\sim\crcr}}}

\begin{document}

\preprint{}

\title{
Precise measurement of $\alpha_K$ and $\alpha_T$ for the 109.3-keV $M$4 transition in $^{125}$Te: Test of internal-conversion theory
}

\author{N. Nica}
\email{nica@comp.tamu.edu}

\author{J.C. Hardy}
\email{hardy@comp.tamu.edu}

\author{V.E. Iacob}

\author{T.A. Werke}

\author{C.M. Folden III}

\author{K. Ofodile}
\altaffiliation {REU summer student from Northern Illinois University, DeKalb, IL 60115}

\affiliation{ Cyclotron Institute, Texas A\&M University, College Station, Texas 77843, USA}
\homepage{http://cyclotron.tamu.edu/}

\author{M.B. Trzhaskovskaya}
\affiliation{Petersburg Nuclear Physics Institute, Gatchina 188300, Russia}

\date{\today}

\begin{abstract}
We have measured the $K$-shell and total internal conversion coefficients (ICCs), $\alpha_K$ and $\alpha_T$, for the 109.3-keV
$M$4 transition in $^{125}$Te to be 185.0(40) and 350.0(38), respectively.  Previously this transition's ICCs were considered
anomalous, with $\alpha$ values lying below calculated values. When compared with Dirac-Fock calculations, our new results show good
agreement.  The $\alpha_K$ result agrees well with the version of the theory that takes account of the $K$-shell atomic
vacancy and disagrees with the one that does not.  This is consistent with our conclusion drawn from a series of high multipolarity
transitions.
 
\end{abstract}

\maketitle

\section{\label{sec:introd} INTRODUCTION}

This study of the 109.3-keV $M$4 transition in $^{125}$Te presents the eighth in a series of $\alpha_K$ measurements
\cite{Ni04,Ni05,Ni07,Ni08,Ni09,Ni14,Ha14,Ni16,Ni17} we began in 2004.  Our goal throughout has been to test the accuracy of calculated $K$-shell
Internal Conversion Coefficients (ICCs) for $E$3 and $M$4 transitions with a precision of $\pm$2\% or better.  We particularly
sought to distinguish between two versions of the theory, one that ignored the atomic vacancy left behind by the emitted
electron, and another that took the vacancy into account.  Prior to 2004, there were very few $\alpha_K$ values known to high
precision, so the treatment of the vacancy and the consequent accuracy of the calculated ICCs were controversial topics \cite{Ra02}. 
 
Today, with our new result there are now eleven $\alpha_K$ values for $E$3 and $M$4 transitions known to better than $\pm$2\%, all
but three being from our work. They cover the range $48 \leq Z \leq 78$ and, so far, they strongly support the ICC model
that includes provision for the atomic vacancy.

What makes such precise measurements possible for us is our having an HPGe detector whose relative efficiency is known to
$\pm$0.15\% ($\pm$0.20\% absolute) over a wide range of energies: See, for example, Ref.\,\cite{He03}.  By detecting both the
$K$ x rays and the $\gamma$ ray from a transition of interest in the same well-calibrated detector at the same time, we can
avoid many sources of error.

The 109.3-keV $M$4 transition in $^{125}$Te is interesting for two reasons.  First, the difference in calculated $\alpha_K$ values
between models that do and do not include the vacancy is 3.4\%, a small but experimentally discernible amount; and second, previous measurements
\cite{Bo52,Re77,Mu82,Sa98a, Sa98b} have consistently produced results that were significantly lower than both model calculations.  The measured $\alpha_T$ values
have been more scattered but also tended to be low \cite{So77,Sa98b,Mu82}.  Of all these published measurements, the first appeared in 1952 and none
is more recent than 1998, so it is reasonable to ask if these ICCs in $^{125}$Te are really anomalous or simply victims of past experimental limitations.

\section {\label{overview} Measurement Overview}

We have described our measurement techniques in detail in previous publications \cite{Ni04,Ni07}
so only a summary will be given here.  If a decay scheme is dominated by a single transition
that can convert in the atomic $K$ shell, and a spectrum of $K$ x rays and $\gamma$ rays is recorded
for its decay, then the $K$-shell internal conversion coefficient for that transition is given by
\begin{equation}
\alpha_K \omega_K = \frac{N_K}{N_\gamma} \cdot \frac{\epsilon_\gamma}{\epsilon_K},
\label{alpha}
\end{equation}
where $\omega_K$ is the fluorescence yield; $N_K$ and $N_{\gamma}$ are the total numbers of observed
$K$ x rays and $\gamma$ rays, respectively; and $\epsilon_K$ and $\epsilon_\gamma$ are the
corresponding photopeak detection efficiencies. As in our recent measurement of a transition in $^{127}$Te \cite{Ni17}, we use the
value $\omega_K$ = 0.875(4) from a systematic evaluation \cite{Sc96}.

\begin{figure}[t]
\epsfig{file=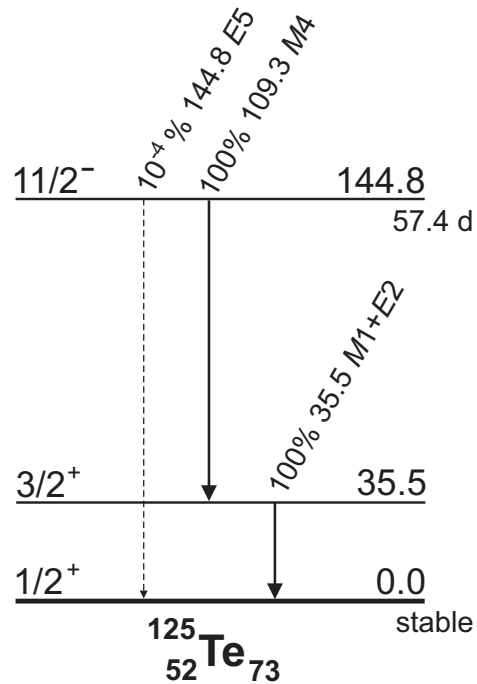,height=9.0cm}
\caption{Decay scheme for the 57-day isomer in $^{125}$Te, illustrating the channels important to this measurement.  The data are taken from Ref.\,\cite{Ka11}.}
\label{fig1}
\end{figure}

The decay scheme of the 57.4-day isomer in $^{125}$Te is shown in Fig.\,\ref{fig1}.  It does not have a single dominant transition
but rather a cascade of two, both of which convert in the $K$ shell and contribute to $N_K$.
To extract an $\alpha_K$ value for the 109.3-keV $M$4 transition
of interest we use a modified version \cite{Ni16} of Eq.\,(\ref{alpha}):
\begin{equation}
\alpha_{K109} = \frac{N_K}{N_{\gamma109}} \cdot \frac{\epsilon_{\gamma109}}{\epsilon_K} \cdot \frac{1}{\omega_K}-\alpha_{K36}
    \cdot \frac{N_{\gamma36}}{N_{\gamma109}} \cdot \frac{\epsilon_{\gamma109}}{\epsilon_{\gamma36}},
\label{alpha2}
\end{equation}
where the subscripts 109 and 36 on a quantity denote the transition -- either the 109.3-keV or 35.5-keV one -- to which
that quantity applies.  Note that the result we are seeking for $\alpha_{K109}$ now depends on $\alpha_{K36}$.

To make the evaluation of uncertainties more transparent, it is convenient to recast this equation in the following form:
\begin{equation}
\alpha_{K109} = \frac{1}{N_{\gamma109}} \cdot \frac{\epsilon_{\gamma109}}{\epsilon_K} \cdot \frac{1}{\omega_K} \cdot N_{K109},
\label{alpha3}
\end{equation}
where
\begin{equation}
N_{K109} = N_K-N_{K36},
\label{alpha4}
\end{equation}
and
\begin{equation}
N_{K36} = \alpha_{K36} \cdot  N_{\gamma36} \cdot \frac{\epsilon_K}{\epsilon_{\gamma36}} \cdot \omega_K .
\label{alpha5}
\end{equation}
Here $N_{K36}$ and $N_{K109}$ represent the contributions to the total $K$ x-rays, $N_K$, due to the 35.5- and 109.3-keV
transitions, respectively.   In this particular case, both contributions are similar in magnitude, so the precision achievable
for $\alpha_{K109}$ suffers as a result.

There is an advantage to having a cascade though: It allows the determination of $\alpha_{T109}$ via the equation,
\begin{equation}
(1+\alpha_{T109}) \cdot \frac{N_{\gamma109}}{\epsilon_{\gamma109}} = (1+\alpha_{T36}) \cdot \frac{N_{\gamma36}}{\epsilon_{\gamma36}}.
\label{alphatot}
\end{equation}
Since both $\alpha_T$ values are much greater than 1, the result extracted for $\alpha_{T109}$ depends directly on the value assumed
for $\alpha_{T36}$.

In our experiment, the HPGe detector we used to observe both $\gamma$ rays and $K$ x rays has been meticulously calibrated
\cite{Ha02,He03,He04} for efficiency to sub-percent precision, originally over an energy range from 50 to 3500 keV but
more recently extended \cite{Ni14} with $\pm$1\% precision down to 22.6 keV, the average energy of silver $K$ x rays.  Over this whole
energy region, precise measured data were combined with Monte Carlo calculations from the CYLTRAN code \cite{Ha92} to yield a very precise
and accurate detector efficiency curve.  In our present study, the $\gamma$ ray of interest at 109.3 keV is well within the energy region for
which our efficiencies are known to a relative precision of $\pm$0.15\%.  The 35.5-keV $\gamma$ ray and the tellurium $K$ x rays, which are
between 27 and 32 keV, all lie comfortably within our extended region of calibration, so the detector efficiency for them can be quoted
to a precision of $\pm$1\% relative to the 109.3-keV $\gamma$ ray.  

\section{\label{exp} Experiment}

\subsection{\label{sprep} Source Preparation}

We obtained tellurium metal powder enriched to 99.93(2)\% in $^{124}$Te from Isoflex USA. With it, we prepared a neutron activation target of $^{124}$TeO by the
molecular plating technique \cite{Pa62, Pa64}. The procedure was in principle identical to the one we used to produce a $^{110}$Cd target for a previous
measurement in this series \cite{Ni16}. A sample of 3.00(2) mg of the $^{124}$Te metal powder was dissolved in 200 $\mu$L of 2 M HNO$_3$ to convert
the metal to its nitrate form. The solution was then evaporated to dryness under a stream of Ar gas. Finally, the sample was
reconstituted with 10 $\mu$L of 0.1 M HNO$_3$ and $\sim$12 mL of pure, anhydrous isopropanol.  This solution was transferred to an electrodeposition cell
\cite{Ma13}, and the $^{124}$Te(NO$_3$)$_4$ was electroplated onto a 10 $\mu$m-thick 99.999\%-pure Al backing (purchased from Goodfellow USA) by
application of +700 V to the Pt anode in the cell. The deposition time was approximately 30 minutes. After deposition, the target was baked at
200$^{\circ}$C under atmospheric conditions for 30 min to ensure the chemical conversion of the thermally unstable $^{124}$Te(NO$_3$)$_4$ into
$^{124}$TeO$_2$. The resulting average thickness of the $^{124}$TeO$_2$ layer was determined to be 308(9) $\mu$g/cm$^2$ as measured by mass.

\begin{figure*}[t]
\epsfig{file=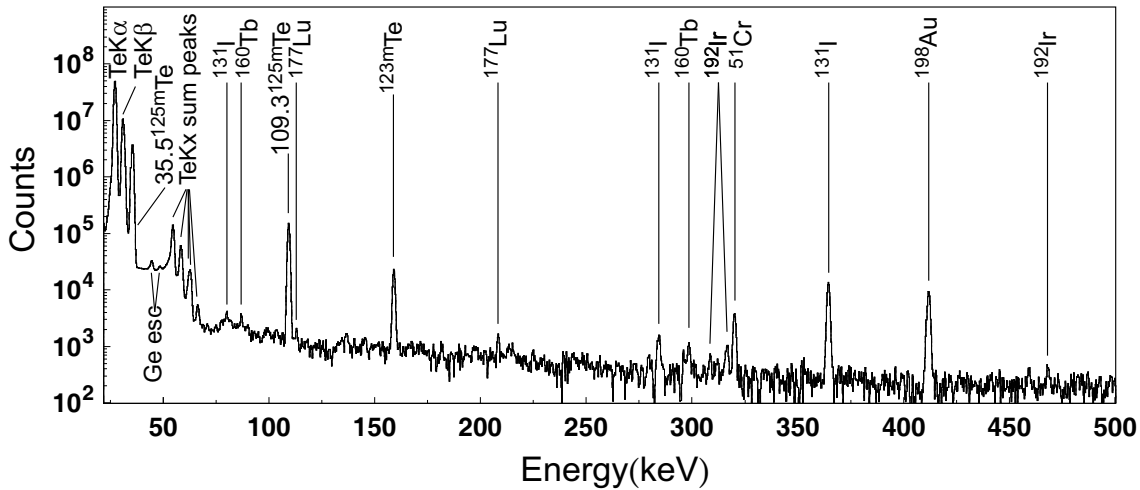,width=15cm}
\caption{Portion of the background-subtracted x- and $\gamma$-ray energy spectrum recorded over a period of 4.7 days, three weeks after activation of enriched
$^{124}$Te.  Peaks are labeled by their $\beta$-decay parent. The cluster of ``sum peaks'' around 60 keV arise from summing of $K$ x rays and $\gamma$ rays from
the 35.5-keV transition with $K$ x rays from the 109-keV transition.  The labeled Ge escape peaks are associated with this cluster.}
\label{fig2}
\end{figure*}

We used identically made $^{\textup{nat}}$TeO$_2$ targets to characterize the product instead of $^{124}$TeO$_2$ because the analysis techniques led to
destruction of the targets. Scanning electron microscopy determined that the TeO$_2$ was mostly uniform, and energy-dispersive X-ray spectrometry (EDS)
verified the elemental composition by an unambiguous identification of Te and O in the sample. Unfortunately, the 1:2 stoichiometric ratio of Te:O could not be
confirmed by the EDS, likely due either to the presence of Al$_2$O$_3$ from the backing or to oxygen-containing compounds present in the carbon-based tape
that was used to secure the sample for analysis. However, the well-known chemistry of Te and the proper visual appearance of the target as a thin layer of
a white solid gave us confidence that the target layer was primarily composed of TeO$_2$.

The electroplated sample was activated for a total of 24 hours in a neutron flux of $\sim7.5\times10^{12}\,n$/(cm$^2$\,s) at the 1-MW TRIGA reactor
in the Texas A\&M Nuclear Science Center.  After removal from the reactor, the sample was stored for 3 weeks and then conveyed to our measurement
location. At that time, the activity from $^{125m}$Te was determined to be $\sim$60 kBq.

\subsection{\label{decaymeas} Radioactive decay measurements}

We acquired spectra with our precisely calibrated HPGe detector and with the same electronics used in its calibration \cite{He03}.  Our
analog-to-digital converter was an Ortec TRUMP$^{TM}$-8k/2k card controlled by MAESTRO$^{TM}$ software.  We acquired 8k-channel spectra at a
source-to-detector distance of 151~mm, the distance at which our calibration is well established.  Each spectrum covered the energy interval
10-2000 keV with a dispersion of about 0.25 keV/channel.
 
After energy-calibrating our system with a $^{152}$Eu source, we recorded sequential $\sim$12-hour decay spectra from the tellurium sample for
a total of 112 hours.  Then, for the following 167 hours we recorded sequential room-background spectra.

\section{\label{sec:analysis} Analysis}
\subsection{\label{subsec:peakfit} Peak fitting}

We summed the spectra recorded from the tellurium sample, and summed the background spectra.  The latter sum was then normalized to the same live
time as the former and was subtracted from it.  A portion of the background-subtracted spectrum recorded from the tellurium source is
presented in Fig.\,\ref{fig2}: It includes the x- and $\gamma$-ray peaks of interest from the decay of $^{125m}$Te, as well as a number of peaks
from contaminant activities.  

In our analysis of the data, we followed the same methodology as we did with previous source measurements \cite{Ni04,Ni05,Ni07,Ni08,Ni09,Ni14,Ha14, Ni16, Ni17}.
We first extracted areas, for essentially all the  x- and $\gamma$-ray peaks in the background-subtracted spectrum.  Our procedure was to determine
the areas with GF3, the least-squares peak-fitting program in the RADWARE series \cite{Rapc}.  In doing so, we used the same fitting procedures as
were used in the original detector-efficiency calibration \cite{Ha02,He03,He04}.

\begin{table}[b]
\caption{\label{table1} The contributions of identified impurities to the energy region of the tellurium $K$ x-ray peaks.  Contributions from two other
impurities -- $^{110m}$Ag and $^{124}$Sb -- were observed at the parts-per-billion level. }
\vspace{2mm}
\begin{ruledtabular}
\begin{tabular}{lll}
 & & Contaminant \\
Source & Contaminant &  contribution (\%) \\[1mm]
\hline \\[-2mm]
& &   \\[-4mm]
~~~$^{121}$Te & Sb $K$ x rays & 0.00204(10) \\
~~~$^{123m}$Te & Te $K$ x rays & 0.0249(6) \\
~~~$^{131}$I & Xe $K$ x rays & 0.00330(8)   \\
\end{tabular}
\end{ruledtabular}
\end{table}
 
\subsection{\label{subsec:imp} Impurities}

Once the areas (and energies) of peaks had been established, we could identify all impurities in the $^{125m}$Te spectrum and carefully check
to see if any were known to produce x or $\gamma$ rays that might interfere with the tellurium $K$ x rays or either of the two $\gamma$-ray peaks
of interest, at 35.5 and 109.3 keV.  As is evident from Fig.\,\ref{fig2}, even the weakest peaks were identified.  In all, we found 3 weak
activities that make a very minor contribution to the tellurium x-ray region; these are listed in Table\,\ref{table1}, where the contributions are
given as percentages of the total tellurium x rays recorded.  No impurities interfere in any way with either of the $\gamma$-ray peaks.

\begin{figure}[b]
\epsfig{file=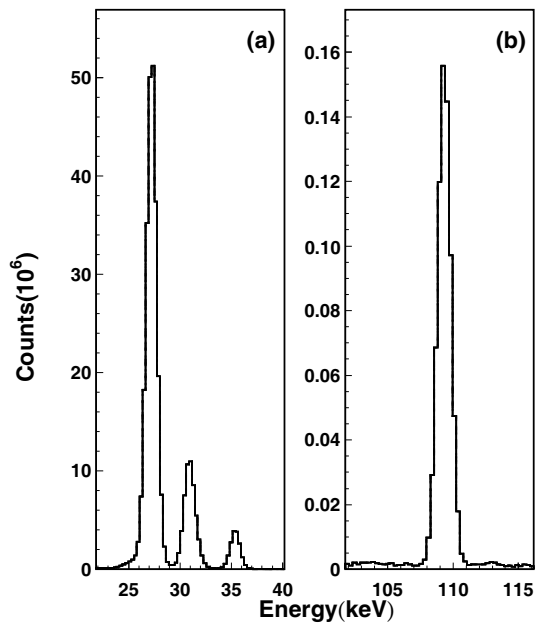,width=7.0cm}
\caption{Spectra for the two energy regions of interest in this measurement, the one on the left including the tellurium $K$ x rays and the one
on the right, the $\gamma$-ray peak at 109.3 keV. These correspond to the full spectrum presented in Fig.\,\ref{fig2}}
\label{fig3}
\end{figure}

Figure \ref{fig3} shows expanded versions of the two energy regions of interest for this measurement: one encompassing the tellurium $K$ x
rays together with the 35.5-keV $\gamma$ ray; and the other, the $\gamma$ ray at 109.3 keV.  In all cases, the peaks lie cleanly on a flat background.
The count totals for the combined $K$ x-ray peaks and for the two $\gamma$-ray peaks at 35.5 and 109.3 keV all appear in Table\,\ref{table2}.  The
impurity total for the combined x-ray peaks appears immediately below their count total; it corresponds to the percentage breakdowns given in Table\,\ref{table1}.

\subsection{\label{subsec:cont} Contamination from the 35.5-keV peak} 

The detector response to 35.5-keV photons adds a significant number of counts to the energy region around the $K$ x-rays. We have previously studied
and discussed at length \cite{Ni07} the scattering tail that extends for over 4 keV towards lower energy from a photon peak at this energy in our detector.  
At our resolution, this tail extends well into the region we integrate to determine the total number of x-ray counts.  Furthermore, each peak in this
energy region is accompanied by two escape peaks arising from the escape of germanium x rays from the detector; these too lie squarely within the x-ray
region.  Based on our earlier scattering studies \cite{Ni07} and on the measured escape-peak ratios for our detector \cite{He03}, we determine the total
contamination of the x-ray region from the 35.5-keV peak to be 8.0(13)\% of the total 35.5-keV peak intensity.  The corresponding number of counts appears
in the first block of Table\,\ref{table2}.

\subsection{\label{subsec:eff} Efficiency ratios} 

In what follows we do not analyze separately the $K_{\alpha}$ and $K_{\beta}$ x rays.  Scattering effects are quite pronounced at these energies and they
are difficult to account for with an HPGe detector when peaks are close together, so we have chosen as before to use only the sum of the $K_{\alpha}$
and $K_{\beta}$ x-ray peaks.  For calibration purposes, we consider the sum to be located at the intensity-weighted average energy of the component
peaks\footnote[1]{To establish the weighting, we used the intensities of the individual x-ray components from Table 7a in Ref. \cite{Fi96}.}---28.03
keV for tellurium.

In order to determine $\alpha_K$ for the 109.3-keV $M$4 transition in $^{125}$Te, we require the efficiency ratio, $\epsilon_{\gamma 109}/\epsilon_K$,
which appears in Eq.\,(\ref{alpha2}).  Following the same procedure as the one we used in analyzing the decay of $^{119m}$Sn \cite{Ni14}, we
employ as low-energy calibration the well-known decay of $^{109}$Cd, which emits 88.0-keV $\gamma$ rays and silver $K$ x rays at a weighted average
energy of 22.57 keV.  Both are relatively close in energy to the respective $\gamma$ and x rays observed in the current measurement.  

\begin{table}[t]
\caption{\label{table2}Corrections to the $^{125}$Te $K$ x rays as well as the 35.5- and 109.3-keV $\gamma$ rays.  Also included is additional
information required to extract a value for $\alpha_K$. }
\vspace{2mm}
\begin{ruledtabular}
\begin{tabular}{lll}
Quantity   &  Value  & Source  \\
\hline \\[-2mm]
\multicolumn{3}{l}{Te ($K_{\alpha} + K_{\beta}$) x rays}  \\
~~ Total counts & 2.9136(27) $\times 10^8$ & Sec.~\ref{subsec:peakfit}  \\
~~ Impurities  & -8.81(18)$\times 10^4$  & Sec.~\ref{subsec:imp}  \\
~~ 35.5-keV peak contamination  & -1.42(22)$\times 10^6$   &  Sec.~\ref{subsec:cont} \\
~~ Net corrected counts, $N_{K}$  & 2.8985(35)$\times 10^8$  &  \\
\hline \\[-2mm]
\multicolumn{3}{l}{Efficiency ratios (including source attenuation)}  \\ 
$a.$ $\epsilon_{\gamma 88}$/$\epsilon_{K23}$ & 1.069(8) & \cite{Ni14} \\
~~ $\epsilon_{K23}$/$\epsilon_{K28}$ & 0.926(5) & \cite{He03,Ch05} \\
~~ $\epsilon_{\gamma109}$/$\epsilon_{\gamma88}$ & 0.9695(15) & \cite{He03,Ch05} \\
~~ $\epsilon_{\gamma109}$/$\epsilon_{K28}$ & 0.960(9) & \\
$b$. $\epsilon_{\gamma88}$/$\epsilon_{\gamma36}$ & 1.002(5) & \cite{He03,Ch05} \\
~~ $\epsilon_{K28}$/$\epsilon_{\gamma 36}$ & 1.012(10) & \\
\hline \\[-2mm]
\multicolumn{3}{l}{35.5-keV $\gamma$ ray}  \\
~~ Total counts, $N_{\gamma 36}$ & 1.6923(13)$\times 10^7$ & Sec.~\ref{subsec:peakfit}  \\
~~ Contribution to x ray, $N_{K36}$ & 1.746(20)$\times 10^8$ & Eq.\,(\ref{alpha5}) \\
\hline \\[-2mm]
\multicolumn{3}{l}{109.3-keV $\gamma$ ray}  \\
~~ Total counts, $N_{\gamma 109}$ & 6.842(11)$\times 10^5$ & Sec.~\ref{subsec:peakfit}  \\
~~ Contribution to x ray, $N_{K109}$ & 1.153(20)$\times 10^8$ & Eq.\,(\ref{alpha4})  \\
\hline \\[-2mm]
\multicolumn{3}{l}{Evaluation of $\alpha_K$}  \\
~~ $N_{K109}/N_{\gamma\,109}$  &  168.6(30)  & This table  \\
~~ Lorentzian correction  &  +0.12(2)\%  &  Sec.\,\ref{subsec:Lor} \\
~~ $\omega_K$  &  0.875(4)  &  \cite{Sc96}  \\
~~ $\alpha_K$ for 109.3-keV transition  & 185.0(40)   &  Eq.\,(\ref{alpha3}) \\
\vspace{-10.pt}
\end{tabular}
\end{ruledtabular}
\end{table}

In our past publications we separately accounted for detector efficiency and attenuation in the source, applying the latter only at the final
derivation of the ICC.  In the $^{125}$Te case, the important contribution of the 35.5-keV $\gamma$ ray makes it necessary for us to incorporate
the source attenuation into all the efficiencies.  Thus, all calculated efficiencies, $\epsilon$, in what follows combine the CYLTRAN computed result \cite{He03} with the
source attenuation obtained from standard tables of attenuation coefficients \cite{Ch05}.

If we now designate the efficiencies (including source attenuation) for the $K$ x rays of tellurium and iodine by $\epsilon_{K28}$ and $\epsilon_{K23}$, respectively, we
can obtain the required ratio, $\epsilon_{\gamma 109}/\epsilon_{K28}$ from the following relation:
\begin{equation}
\frac{\epsilon_{\gamma 109}}{\epsilon_{K28}} = \frac{\epsilon_{\gamma88}}{\epsilon_{K23}} \cdot
\frac{\epsilon_{K23}}{\epsilon_{K28}} \cdot \frac{\epsilon_{\gamma 109}}{\epsilon_{\gamma 88}}.
\label{effratio1}
\end{equation} 
We take the $^{109}$Cd ratio $\epsilon_{\gamma 88}/\epsilon_{K23}$ from our previously reported measurement \cite{Ni14}.  The ratio
$\epsilon_{\gamma 109}/\epsilon_{\gamma 88}$ is close to unity and determined with high precision from our known detector efficiency curve
calculated with the CYLTRAN code \cite{He03}, while $\epsilon_{K23}/\epsilon_{K28}$ comes from a CYLTRAN calculation as well but in an energy region
with higher relative uncertainty.  Nevertheless, the energy span is not large so the uncertainty is only $\pm$0.5\%.  The values of all four efficiency
ratios from Eq.\,(\ref{effratio1}) appear in part $a$ of the second block of Table \ref{table2}.

In evaluating Eq.\,(\ref{alpha5}), we also require the efficiency ratio $\epsilon_{K28}/\epsilon_{\gamma 36}$, which can be expressed as follows:
\begin{equation}
\frac{\epsilon_{K28}}{\epsilon_{\gamma 36}} = \frac{\epsilon_{K23}}{\epsilon_{\gamma88}} \cdot
\frac{\epsilon_{K28}}{\epsilon_{K23}} \cdot \frac{\epsilon_{\gamma 88}}{\epsilon_{\gamma 36}}.
\label{effratio2}
\end{equation} 
Here the first terms on the right are the same as the corresponding terms in Eq.\,(\ref{effratio1}) except that they are inverted.  The third term,
$\epsilon_{\gamma 88}/\epsilon_{\gamma 36}$, which comes from a CYLTRAN calculation, appears in part $b$ of the second block of Table \ref{table2}
together with the result for $\epsilon_{K28}/\epsilon_{\gamma 36}$.

\subsection{\label{subsec:Lor} Lorentzian correction}

As explained in our previous papers (see, for example, Ref.\,\cite{Ni04}) we use a special modification of the GF3 program
that allows us to sum the total counts above background within selected energy limits.  To account for possible missed counts outside those limits, the
program adds an extrapolated Gaussian tail.  This extrapolated tail does not do full justice to x-ray peaks, whose Lorentzian shapes reflect the finite
widths of the atomic levels responsible for them.  To correct for this effect we compute simulated spectra using realistic Voigt functions to generate
the x-ray peaks, and we then analyze them with GF3, following exactly the same fitting procedure as is used for the real data, to ascertain how
much was missed by this approach.  The resultant correction factor appears as a percent in the fifth block of Table\,\ref{table2}.

\section{\label{sec:results} Results and Discussion}

With one exception, all the quantities required to evaluate Eqs.\,(\ref{alpha3}-\ref{alpha5}) are available in Table\,\ref{table2}.  The exception becomes
evident when we seek to use Eq.\,(\ref{alpha5}) to derive $N_{K36}$, the contribution of the 35.5-keV transition to the $K$ x-rays: We need to calculate the
$K$-shell ICC for the 35.5-keV transition, $\alpha_{K36}$.  This is a mixed $M$1 and $E$2 transition with a measured mixing ratio of $\delta$ = 0.031(3)
\cite{Ka11}. Our ICC calculations are made within the Dirac-Fock framework with the option either to ignore the $K$-shell vacancy or to include it in the
``frozen-orbital" approximation \cite{Ba02}.  Taking the transition energy to be 35.4925(5) keV \cite{Ka11}, we find that the two different calculations yield
values of $\alpha_{K36}$ that differ by less than 1\%: 11.61 (no vacancy) and 11.69 (vacancy included).  So as not to prejudice our result for the 109.3-keV
transition, we adopt the value 11.65(4), which encompasses both possibilities.  Substituting this value into Eq.\,(\ref{alpha5}) we obtain the $N_{K36}$
result that appears in the third block of the table.

Next, using the corrected number of counts in the $K$ x-ray peaks, $N_K$, which is given on the last line of the first block in the table, we obtain $N_{K109}$
from Eq.\,(\ref{alpha4}); that result is given in the fourth block of Table\,\ref{table2}.  Finally, after applying the Lorentzian correction to $N_{K109}$ we
use Eq.\,(\ref{alpha3}) to derive the result:
\begin{equation}
\alpha_{K109} = 185.0(40),
\label{alphaK}
\end{equation}
where the uncertainty is dominated by contributions from the efficiency ratios and $\omega_K$.

Making use of Eq.\,(\ref{alphatot}), we can relate the total ICCs for the 35.5- and 109.3-keV transitions with the following relation:
\begin{equation}
\alpha_{T109} = 23.95(25) ~(1+\alpha_{T36}) - 1,
\label{alphatotrel}
\end{equation}
where we have used the ratio $\epsilon_{\gamma 109}/\epsilon_{\gamma 36}$ = 0.971(10) based on our known detector efficiency response \cite{He03}, and we have
included a 0.31\% correction to account for real coincidence summing.  Since the amount of summing with $K$ x rays is different for the two $\gamma$ rays, the
effect needs to be incorporated into the derivation of $\alpha_{T109}$, which involves a $\gamma$-ray ratio.  The effects cancel out when the ratios are of x
rays to $\gamma$ rays for an individual transition, as in the derivation of $\alpha_{K109}$.

To obtain $\alpha_{T109}$ from Eq.\,(\ref{alphatotrel}) we need to calculate a value for the total ICC for the 35.5-keV transition. If the atomic vacancy is
ignored, the calculated value of $\alpha_{T36}$ is 13.61; if the vacancy is included, the value is 13.70.  Once again we choose the average with an assigned
uncertainty that encompasses both values, 13.66(5).  Substituting this value into Eq.\,(\ref{alphatotrel}), we obtain
\begin{equation}
\alpha_{T109} = 350.0(38).
\label{alphaT}
\end{equation}
Here the uncertainty is overwhelmingly due to the contribution from the efficiency ratio.

Both $\alpha_{K109}$ and $\alpha_{T109}$ have been measured a number of times in the past.  Previous results for $\alpha_{K109}$ are 159(24) \cite{Bo52},
151(11) \cite{Re77}, 169(7) \cite{Mu82} and 166(9) \cite{Sa98a,Sa98b}\footnote[2]{Essentially the same result appears in two references with the same
authors; yet neither paper references the other.  We assume here that there was only one measurement.}.  The first of these results, published in 1952, is statistically consistent
with ours but the three more recent ones, appearing between 1977 and 1998, are lower by two or more of their standard deviations.  In the case of $\alpha_{T109}$, the
previous results are 357(11) \cite{So77}, 304(17) \cite{Mu82} and 318(40) \cite{Sa98b}.  Once again, the earliest measurement, from 1977, is consistent
with our result, as is the most recent 1998 result.  The 1982 measurement is low by more than two of its standard deviations.  Although overall there is
some agreement with our results, all but one of the previous measurements has been low, and the averages have led to the conclusion that the ICCs for this
transition are anomalously low.  Our measurements show this to be false.
  
We compare our results with three different theoretical calculations in Table\,\ref{table3}. All three calculations were made within
the Dirac-Fock framework, but one ignores the presence of the $K$-shell vacancy while the other two include it using different approximations: the
frozen-orbital approximation, in which it is assumed that the atomic orbitals have no time to rearrange after the electron's removal; and the SCF approximation, in
which the final-state continuum wave function is calculated in the self-consistent field (SCF) of the ion, assuming full relaxation of the ion orbitals. 

\begin{table}[t]
\caption{\label{table3}Comparison of the measured $\alpha_K$ and $\alpha_T$ values for the 109.276(15)-keV $M$4 transition from $^{125m}$Te with
calculated values based on three different theoretical models, one that ignores the $K$-shell vacancy and two that deal with it either in the ``frozen-orbital" (FO)
approximation or the self-consistent field (SCF) approximation (see text).  The uncertainties on the calculations reflect the uncertainty in the measured transition
energy.  Shown also are the percentage deviations, $\Delta$, from the experimental value calculated as (experiment-theory)/theory.  For a full description of the various
models used to determine the conversion coefficients, see Ref.\,\cite{Ni04}.}
\vspace{2mm}
\begin{ruledtabular}
\begin{tabular}{lllll}
\multicolumn{1}{l}{Model}  & \multicolumn{1}{c}{~~$\alpha_K$} & \multicolumn{1}{c}{~~$\Delta$(\%)} & \multicolumn{1}{c}{~~$\alpha_T$} & \multicolumn{1}{c}{~~$\Delta$(\%)} \\
\hline \\[-3mm]
Experiment & 185.0(40)  & & 350.0(38) &  \\
Theory: & & \\
~~~No vacancy  & 179.5(1) & +3.0(22) & 348.7(3) & $+0.4(11)$ \\
~~~Vacancy, FO  & 185.2(1) & $-0.1(22)$ & 355.6(3) & $-1.6(11)$ \\
~~~Vacancy, SCF  & 184.2(1) & $ +0.4(22)$ & 354.2(3) & $-1.2(11)$ \\
\vspace{-10.pt}
\end{tabular}
\end{ruledtabular}
\end{table}

The percentage deviations given for $\alpha_K$ in Table\,\ref{table3} indicate excellent agreement between our measured result and the two calculations that include some
provision for the atomic vacancy.  Our measurement disagrees by 1.4 standard deviations with the calculation that ignores the vacancy.  This outcome is barely significant
statistically but it is consistent with our previous seven precise $\alpha_K$ measurements on $E$3 and $M$4 transitions in $^{111}$Cd \cite{Ni16}, $^{119}$Sn \cite{Ni14,Ha14},
$^{127}$Te \cite{Ni17}, $^{134}$Cs \cite{Ni07,Ni08}, $^{137}$Ba \cite{Ni07,Ni08}, $^{193}$Ir \cite{Ni04,Ni05} and $^{197}$Pt \cite{Ni09}, all of which agreed well with
calculations that included the vacancy, and disagreed -- some by many standard deviations -- with the no-vacancy calculations.   

The situation is more ambiguous for $\alpha_T$: There our measured result agrees best with the no-vacancy calculation but it is consistent as well with the SCF version of the
calculation, which includes the vacancy. Note also that the measured value of $\alpha_{T109}$ depends on a calculated value for $\alpha_{T36}$, which in turn
depends on the measured $E$2/$M$1 mixing ratio \cite{Ka11} for the 35.5-keV transition.  If that mixing ratio were wrong, it could have an impact on our $\alpha_{T109}$ result.

\section{Conclusions}

Our measurements of the $K$-shell and total internal conversion coefficients for the 109.3-keV $M$4 transition from $^{125m}$Te has yielded values that are no
longer anomalous when compared with calculations that use the Dirac-Fock theory.  In addition, the result for $\alpha_{K109}$ is precise enough to show a
statistical preference, albeit small, for one particular version of the Dirac-Fock theory: It agrees well with the version that includes the atomic vacancy and disagrees
(by $\sim$1.4$\sigma$) with theory if the vacancy is ignored.  We have now made eight precise $\alpha_K$ measurements for $E$3 and $M$4 transitions in nuclei with a wide
range of $Z$ values.  Their corresponding conversion-electron energies also ranged widely, from $\sim$4 keV in $^{193}$Ir to $\sim$630 keV in $^{137}$Ba.  These
measurements together present a consistent pattern that supports the Dirac-Fock theory for calculating $K$-shell internal conversion coefficients provided that it takes
account of the atomic vacancy.

Early results from our program influenced a 2008 reevaluation of ICCs by Kib\'{e}di $et~al.$ \cite{Ki08}, who also developed BrIcc, a new data-base obtained from the basic
code by Band $et~al.$ \cite{Ba02}.  In conformity with our conclusions, BrIcc employed a version of the code that incorporates the vacancy in the frozen-orbital
approximation.  The BrIcc data-base has been adopted by the National Nuclear Data Center (NNDC) and is available on-line for the determination of ICCs.  Our experimental
results obtained since 2008 continue to support that decision.

Though we have obtained a $\pm$1.1\% result for the total ICC of the 109.3-keV transition, it is still not precise enough to allow any conclusions to be drawn concerning
a preferred version of the Dirac-Fock theory for $\alpha_{T109}$.  The calculated results differ from one another by less than 2\%, and our result has statistical overlap
with both the no-vacancy and SCF vacancy-inclusive versions.  Any definitive conclusion must await a measurement with even greater precision.   Certainly, though, we can
already conclude that the large discrepancy with theory, suggested by previous measurements, can be ruled out.

\begin{acknowledgments}

We thank the Texas A\&M Nuclear Science Center staff for their help with the neutron
activations.  This material is based upon work supported by the U.S. Department of Energy, Office of Science,
Office of Nuclear Physics, under Award Number DE-FG03-93ER40773, and by the Robert A. Welch Foundation under
Grant No.\,A-1397.

\end{acknowledgments}

\end{document}